\documentclass{article}
\usepackage[T1]{fontenc} 
\usepackage[utf8]{inputenc} 
\usepackage{ismir,amsmath,cite,url}
\usepackage{graphicx}
\usepackage{booktabs}
\usepackage{color}
\usepackage{dblfloatfix} 

\usepackage[firstpage]{draftwatermark}
\definecolor{lightgray}{rgb}{0.9,0.9,0.9}
\definecolor{darkgray}{rgb}{0.4,0.4,0.4}
\SetWatermarkFontSize{12pt}
\SetWatermarkScale{1.1}
\SetWatermarkAngle{90}
\SetWatermarkHorCenter{202mm}
\SetWatermarkVerCenter{170mm}
\SetWatermarkColor{darkgray}
\SetWatermarkText{Late-Breaking / Demo Session Extended Abstract, ISMIR 2024 Conference}

\def\lbd{}	

\title{Exploring Tokenization Methods for Multitrack Sheet Music Generation}

\multauthor
{Yashan Wang$^{1,\sharp}$\thanks{$^\sharp$ These authors contributed equally.} \hspace{1cm} Shangda Wu$^{1,\sharp}$ \hspace{1cm} {Xingjian Du}$^{2}$ \hspace{1cm} Maosong Sun$^{1,3,\flat}$\thanks{$^\flat$ Corresponding author.}}
{ $^1$ Central Conservatory of Music, China \\$^2$ University of Rochester, USA $^3$ Tsinghua University, China\\
{\tt\small \{alexis\_wang, shangda\}@mail.ccom.edu.cn, \ x.du@rochester.edu, \ sms@tsinghua.edu.cn} \\
}
\def\authorname{Y. Wang, S. Wu, X. Du, and M. Sun}

\begin{document}

\maketitle
\begin{abstract}
This study explores the tokenization of multitrack sheet music in ABC notation, introducing two methods—bar-stream and line-stream patching. We compare these methods against existing techniques, including bar patching, byte patching, and Byte Pair Encoding (BPE). In terms of both computational efficiency and the musicality of the generated compositions, experimental results show that bar-stream patching performs best overall compared to the others, which makes it a promising tokenization strategy for sheet music generation.

\end{abstract}
\section{Introduction}\label{sec:introduction}

Sheet music generation, particularly using ABC notation—a compact, text-based format, has gained prominence in symbolic music generation.  \cite{sturm2016music, wu2023chord, wu2023tunesformer, qu2024mupt, casini2024investigating}. Tokenizing multitrack ABC notation in language models presents unique challenges due to inter-track dependencies. FolkRNN \cite{sturm2016music} represented musical elements like pitch and duration as multi-character tokens. In contrast, CLaMP \cite{wu2023clamp} and bGPT \cite{wu2024beyond} introduced bar patching and byte patching, respectively, which tokenize score text into patches and then decode them with a character-level decoder. MuPT \cite{qu2024mupt} used the Byte Pair Encoding (BPE) method \cite{10.5555/177910.177914} from NLP. Nevertheless, challenges regarding musicality and computational efficiency still exist.

In this work, we investigate tokenization as a critical initial step in training a sheet music generation model, aiming to minimize computational costs while maintaining the quality of the generated music. Building on bar patching and byte patching methods, we introduce two new techniques—bar-stream patching and line-stream patching. We evaluate all patching methods, including BPE, within a pre-training and fine-tuning framework. 

\section{Methodology}

\subsection{Model Architecture}

We adopted Tunesformer \cite{wu2023tunesformer}, a hierarchical GPT-2 \cite{radford2019language} decoder architecture, for our patching methods. In this framework, patch-level decoders embed and process patches to generate features for a character-level decoder, which performs auto-regressive character prediction. The context lengths are determined by the patch length for the patch-level decoder and the patch size for the character-level decoder. For BPE, we use a standard GPT-2 decoder.

\subsection{Data Tokenization}

To ensure multitrack score voice alignment, we use interleaved ABC notation \cite{qu2024mupt} for multiple musical parts. Then, we tokenize score text with four patching methods and BPE, as shown in Fig. 1. Existing methods include:

\textbf{Bar patching:} Divide score text into bar patches, where each bar corresponds to a single voice, and truncate/pad bars based on patch size.

\textbf{Byte patching:} Divide score text into fixed-length patches regardless of musical score semantics.

\textbf{BPE:} A 50,000-token vocabulary was created through an iterative tokenization approach that merges frequent character or sub-word pairs in the score text.

To avoid the truncation in bar patching and ensure division according to semantic units of musical scores, two patching methods are proposed:

\textbf{Bar-stream patching}: An improvement on bar patching. First, the score text is divided into bars. Then, each bar is split into fixed-length patches as per the patch size; if a bar's final patch is shorter than the patch size, it is padded.

\textbf{Line-stream patching}: Like bar-stream patching, but this method divides the score by line breaks. In interleaved ABC, each line represents a bar with all voices.

\subsection{Dataset}

Pre-training used an in-house 160K ABC-notation score dataset. To evaluate models' generalization with different tokenization, we fine-tuned on three classical music datasets of different instrumentation: 398 Bach chorales \cite{cuthbert2010music21}, 103 Haydn string quartets \cite{gotham2023openscore}, 54 Mozart piano sonatas \cite{hentschel2021annotated}. Additionally, data augmentation on 15 key signatures was done in both pre-training and fine-tuning.

\begin{figure*}[t] 
\centering
\includegraphics[width=0.8\textwidth]{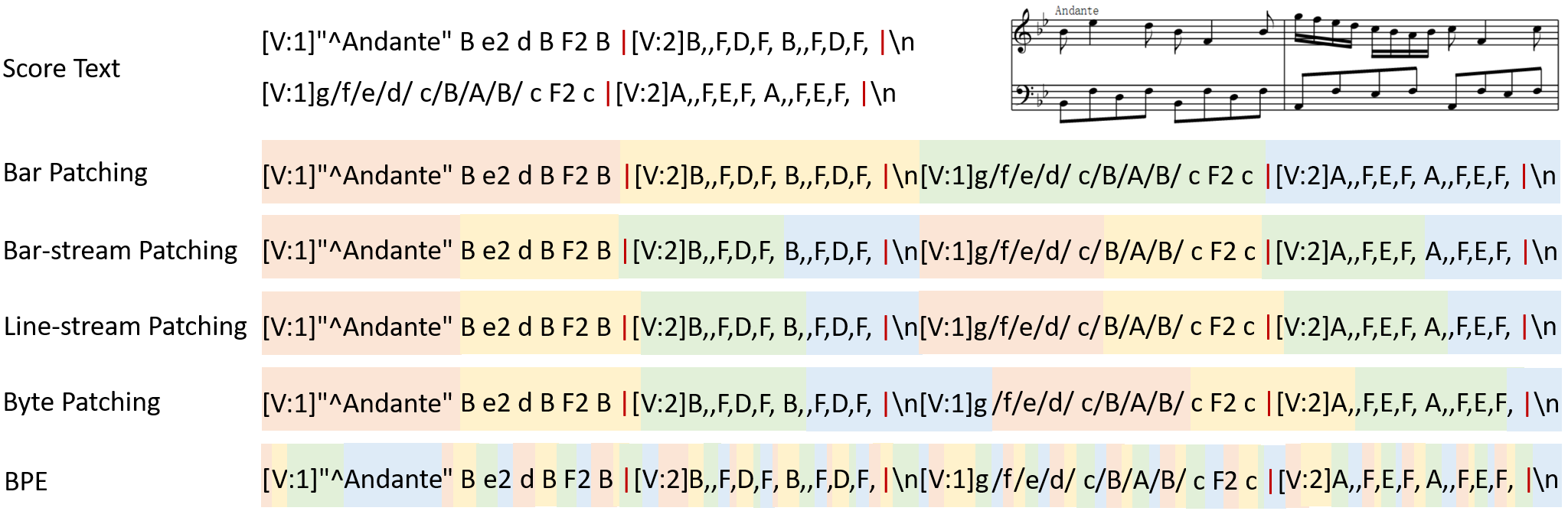} 
\caption{An example of score text and various tokenization implementations with colors marking token boundaries.}
\label{fig:example}
\end{figure*}

\begin{table*}[htb!]
\centering
\resizebox{\textwidth}{!}{%
\begin{tabular}{l|ccccccccccc}
 \midrule
Tokenization & Parameters & Sec/Epoch &  \multicolumn{3}{c}{Inference Speed} & \multicolumn{3}{c}{BPB} & \multicolumn{3}{c}{CLaMP 2 Score} \\ 
 \cmidrule(lr){4-6} \cmidrule(lr){7-9} \cmidrule(lr){10-12}
 & & & Bach & Haydn & Mozart & Bach & Haydn & Mozart & Bach & Haydn & Mozart \\ 
\midrule
Byte patching      & 65,872,896 & \textbf{963} & \textbf{597.1} & \textbf{630.4} & \textbf{623.8} & 0.2795 & 0.3682 & 0.3900 & 0.9767 & 0.9071 & 0.8068 \\
Line-stream patching  & 65,872,896 & 1107 & 549.7 & 564.8 & 569.0 & 0.2772 & 0.3797 & 0.3958 & 0.9734 & 0.8916 & 0.8213 \\
Bar-stream patching   & 65,872,896 & 1063 & 446.3 & 465.6 & 449.6 & 0.2539 & 0.3526 & 0.3879 & 0.9781 & \textbf{0.9228} & \textbf{0.8225} \\
Bar patching       & 70,628,352 & 2848 & 226.1 & 210.9 & 204.3 & \textbf{0.2479} & 0.3515 & 0.3920 & \textbf{0.9813} & 0.9045 & 0.7531 \\
BPE       & 84,074,496 & 4071 & 91.0  & 80.2  & 71.1  & 0.2591 & \textbf{0.3340} & \textbf{0.3542} & 0.9687 & 0.9050 & 0.7005 \\
\bottomrule
\end{tabular}%
}
 \caption{Comparison of evaluation results among different tokenization methods.}
 \label{tab:example}
\end{table*}

\section{Experiments}

\subsection{Settings}

For patching methods, we used a 6-layer patch-level decoder and a 3-layer character-level decoder. For BPE, a 6-layer decoder was directly applied. To balance bar truncation and efficiency, the patch size was set to 64 for bar patching (covering 97.7\% of all bars) and 16 for other patching methods where truncation is not an issue. The patch length was 512 for all patching methods and 4096 for BPE, ensuring comparable score lengths across attention spans. All pre-training was carried out using 2 H800 GPUs with the batch size maximized.

\subsection{Evaluation Metrics}

We evaluated models' efficiency and musicality across different tokenization strategies using these metrics:

\textbf{Sec/Epoch}: This represents the average duration of each pre-training epoch, measured in seconds.

\textbf{Bits-per-byte (BPB)}: Calculates the average bits to predict the next token on the validation set.

\textbf{Inference Speed}: Average characters generated per second during inference.

\textbf{CLaMP 2 Score}: Calculated by extracting semantic features with CLaMP 2 \cite{wu2024clamp2multimodalmusic} and computing the cosine similarity between the validation set and the generated data. A higher score means the generated data is more similar to the real data.

\subsection{Results}

Regarding efficiency, byte patching, line-stream patching, and bar-stream patching require shorter training times and have faster inference speeds, with byte patching performing the best. Bar patching and BPE are less computationally efficient because bar patching has a larger patch size and BPE has a longer context length.

For BPB, BPE generally performs best. This is likely because BPE tokenizes score text into high-frequency combinations, thus providing the model with more prior knowledge compared to the character-level decoding in patching methods.

However, BPE underperforms in CLaMP 2 Scores, suggesting a semantic gap between the generated results and real music. In contrast, bar-stream patching achieves high CLaMP 2 Scores. It not only avoids bar truncation issues but also incorporates prior knowledge of bar units during patching, leading to better musicality.

Overall, our experiments show that bar-stream patching is the top-performing method, presenting a balanced performance across all metrics. It combines high training and inference efficiency with generated results that closely resemble real classical compositions.

\section{Conclusion}

In this study, we explored tokenization methods for sheet music generation based on ABC notation. We introduced bar-stream and line-stream patching and compared them with bar patching, byte patching, and BPE. Focusing on the balance between computational efficiency and musicality, the results demonstrated that bar-stream patching outperformed the others in general.

For future work, we will scale up the model size and dataset with employing bar-stream patching and a hierarchical decoder. Additionally, we will establish a classical-music-centered dataset for fine-tuning to enhance the musicality of the generated results.

\bibliography{ISMIRtemplate}

\end{document}